\documentstyle[fps,twoside,fleqn,espcrc2]{article}
   
   \newcommand{\nc}{\newcommand}
   \nc{\vct}[1]{\vec{#1}\hspace{0.3ex}}
   \nc{\hut}[1]{\rlap{$#1$}\hat{\rule[1.6ex]{0.5em}{0.0ex}}}
   \nc{\ds}{\displaystyle}
   \nc{\be}{\begin{equation}}
   \nc{\eq}{\end{equation}}
   \nc{\bc}{\begin{center}}
   \nc{\ec}{\end{center}}
   \nc{\bea}{\begin{eqnarray}}
   \nc{\eqa}{\end{eqnarray}}
   \nc{\non}{\nonumber}
   \nc{\tlift}{\rule[-0.5ex]{0em}{2.4ex}}
   \nc{\al}{\alpha}
   \nc{\bt}{\beta}
   \nc{\dl}{\delta}
   \nc{\kp}{\kappa}
   \nc{\sg}{\sigma}
   \nc{\Gm}{\Gamma}
   \nc{\lph}{\Phi}                       
   \nc{\PP}{\tilde{\Phi}}                
   \nc{\lra}{\longrightarrow}
   \nc{\lla}{\longleftarrow}
   \nc{\tl}{\tilde}
   \nc{\p}{\partial}
   \nc{\khat}{\widehat{k}}
   \nc{\pll}{\parallel}
   \nc{\hmu}{\hat{\mu}}
   \nc{\Real}{\mbox{\rm Re}\,}
   \nc{\asinh}{\,{\rm asinh}\,}
   \nc{\lag}{\langle}
   \nc{\rag}{\rangle}
   \nc{\ZZ}{\bf Z}
   \nc{\lb}{\label}
\title{Unstable particles in finite volume:\\
       The broken phase of the $4$-d $O(4)$ non-linear
       $\sigma$-model}
\author{
        Frank Zimmermann\address{
        Institut f\"ur Theoretische Physik E, RWTH Aachen,
        W-5100 Aachen, FRG}$\hspace{-0.8ex}^{,\,}$\address{
        HLRZ c/o KFA J\"ulich, P.O.Box 1913,
        W-5170 J\"ulich, FRG}$\hspace{-0.8ex}^{,\,}$\thanks{
        speaker at the conference},
        J\"org Westphalen$^{\rm a}$,
        \\ Meinulf G\"ockeler$^{\rm a,\,}$$^{\rm b,\,}$\thanks{
        supported by the {\em Deutsche Forschungsgemeinschaft}}
        and Hans A.~Kastrup$^{\rm a}$}%
\begin{document}
\begin{abstract}
 According to a proposal of L\"uscher it is possible to determine elastic
scattering phases in infinite volume from the energy spectrum of
two-particle states in a periodic box.
   We demonstrate the applicability of this method in the broken phase of
the 4-dimensional $O(4)$ non-linear $\sg$-model in a Monte-Carlo study on
finite lattices.
   This non-perturbative approach also permits the study of unstable
particles, the $\sg$-particle in our case.
We observe the $\sg$-resonance and extract its mass and its width.
\end{abstract}
\maketitle
\section{INTRODUCTION}
\noindent
Recently L\"uscher derived a relation between the energy spectrum of
two-particle states in a finite box with periodic boundary conditions and
elastic scattering phase shifts defined in infinite volume \cite{lue2part}.
Since two-particle energy levels are calculable by Monte Carlo
techniques, this relation opens the possibility to extract phase shifts
from numerical simulations on finite lattices.
The scattering lengths have already been examined by means of
the asymptotic large-volume behaviour of the two-particle energy
levels \cite{chris}.
Furthermore, L\"uscher's relation for two dimensions  has recently been
used to extract the scattering phase shifts of the $O(3)$ non-linear
$\sg$-model \cite{luewo} and to study resonance scattering for two
coupled Ising systems \cite{gatlan}.
As a next step towards the determination of resonance parameters in
QCD \cite{luereson} we show the applicability of the relation
in the 4-dimensional $O(4)$ non-linear $\sg$-model.
We investigate this model in its broken phase, where the
$\sg$-particle is unstable.

In order to study two-particle scattering in finite (spatial) volume
L\"uscher starts with a non-relativistic model considering two identical
massive bosons in a periodic cubic box. The box size is assumed to be
larger than the interaction range. He shows that the energy spectrum is
determined by the scattering phases at the energy eigenvalues. For states
below the inelastic threshold this result is then carried over to
massive relativistic quantum field theories by invoking an effective
Schr\"odinger equation.

Since this whole reasoning only works in the absence of massless (physical)
particles, we have to equip the Goldstone bosons with a mass before we can
apply L\"uscher's formulas. So we add an
external source term to the lattice action
of the 4-dimensional $O(4)$ non-linear $\sg$-model:
\be
      S = -2 \kp \sum_{x} \sum_{\mu=1}^4
           \Phi^\alpha_x \Phi^\alpha_{x+\hmu}
          + J \sum_{x} \Phi^4_x \,.
\lb{eq:action}
\eq
The scalar field is represented as a four component vector
$\Phi^\alpha_x$ of unit length:
$\Phi^\alpha_x \Phi^\alpha_x \!=\! 1$.
\vspace*{-0.7ex}
\section{THEORETICAL BACKGROUND}
\noindent
Consider a system of two Goldstone bosons (``pions'') of mass $m_\pi$ with
zero total momentum in a cubic box of size $L^3$. The two-particle states
in the elastic region are characterised by centre-of-mass energies $W$
or momenta $\vct{k}$ defined through
\be
       W = 2 \sqrt{m_\pi^2 + \vct{k}^2} \; , \qquad k = |\vct{k}| \,,
\lb{eq:energ}
\eq
with $W$ and $k$ in the ranges:
\be
       2 m_\pi < W < 4 m_\pi \quad \Leftrightarrow \quad
       0 < k/m_\pi < \sqrt{3} \,.
\lb{eq:elast}
\eq
They are classified according to irreducible representations of the cubic
group. Their discrete energy spectrum $W_j$, $j=0,1,2,\ldots,$
is related to the scattering phase shifts $\dl_{l}$ with angular momenta
$l$ which are allowed by the cubic symmetry of the states \cite{lue2part}.

In the subspace of cubically invariant states only angular momenta
$l=0,4,6,\ldots$ contribute. Due to the short-range interaction it seems
reasonable to neglect all $l\!\neq\! 0$. Then an energy value $W_j$
belongs to the two-particle spectrum, if the corresponding momentum
$k_j=\sqrt{(W_j/2)^2-m_\pi^2}$ is a solution of
\be
    \dl_0 (k_j) =  - \phi \left(\frac{k_j L}{2\pi}\right) + j \pi \,.
\lb{eq:scatter}
\eq
The continuous function $\phi(q)$ is given by
\be
    \tan \left(-\phi(q)\right)
          = \frac{\ds q \pi^{3/2}}{\ds {\cal Z}_{00}(1,q^2)}
     \quad, \quad \phi(0) = 0 \,,
\eq
with ${\cal Z}_{00} (1,q^2)$ defined by analytic continuation of
\be
    {\cal Z}_{00} (s,q^2) = \frac{1}{\sqrt{4\pi}}
           \sum_{\vct{n}\in\ZZ^{3}}
              \left(\vct{n}^{2}-q^2\right)^{-s} \,.
\eq
This result holds only for $0\! <\! q^2\! <\! 9$, which is fulfilled in our
simulations, and provided finite volume polarisation effects and scaling
violations are negligible \cite{lue2part}.

In our Monte-Carlo investigation we calculate the momenta $k_j$
corresponding to the two-particle energy spectrum $W_j$ for a
given size $L$, and can then read off the scattering phase $\dl_0(k_j)$
from eq.~(\ref{eq:scatter}).
In order to scan the momentum dependence of the phase shift
we have to vary the spatial extent $L$ of the lattice.
%
\vspace*{-0.7ex}
\section{HOW TO EXTRACT THE TWO-PARTICLE SPECTRUM}
\noindent
The two-particle states in our model are also classified according to the
remaining $O(3)$-symmetry: They have ``isospin'' $0,1,2$. Since we expect
the $\sg$-resonance to be in the isospin-$0$ channel, we restrict
ourselves to that case. The corresponding two-particle operators on a
lattice of spatial extent $L$ are defined according to
\be
    {\cal O}_{i} (t) =
                \sum_{a=1}^{3}
                \sum_{\vct{p}} f_{i} (\vct{p}) \;
                \PP^{a}_{-\vct{p},t} \; \PP^{a}_{\vct{p},t} \,,
                \quad \vct{p} \in \ZZ^{3}_{L}
\lb{eq:2partop}
\eq
where $f_i(\vct{p})$, $i=1,2,\ldots$ is some basis of wave functions and
$\PP^{\al}_{\vct{p},t}$ is the spatial Fourier transform of the lattice
field $\Phi^{\al}_{x}$:
\be
     \PP^{\al}_{\vct{p},t} = L^{-3}
                             \sum_{\vct{x}} \Phi^{\al}_{\vct{x},t} \;
                              e^{2\pi i\vct{x}\cdot\vct{p}/L} \,.
\eq
\hspace*{\parindent} We want to compare  two different kinds of cubically
invariant wave functions. First we take plane waves,
\be
      f_{i} (\vct{p}) = \dl_{\,i,\vct{p}^2} \quad , \quad i=1,2,\ldots
\lb{eq:plawav}
\eq
Our second choice is motivated by the periodic singular solutions of the
Helmholtz equation (see \cite{lue2part}),
\be
      f_{i} (\vct{p}) = (\vct{p}^2-q_{i}^2\,)^{-1}
      \quad , \quad i=1,2,\ldots \;,
\lb{eq:luewav}
\eq
with values $q_i = (L/2\pi)\sqrt{(W_i/2)^2-m_\pi^2}$ using as input
the plane wave energy spectrum $W_i$ of some test runs.
These latter wave functions should have an enhanced projection especially
at energies close to the resonance mass.

We also take into account the $\sg$-field at zero
momentum, since it has the same quantum numbers as ${\cal O}_{i}(t)$
defined in eq.~(\ref{eq:2partop}):
\be
    {\cal O}_{0} (t) = \PP^{4}_{\vct{0},t} = L^{-3}
                \sum_{\vct{x}} \Phi_{\vct{x},t}^{4} \,.
\eq
\hspace*{\parindent} The eigenvalues of the two-particle correlation
function matrix
\be
   C_{ij} (t) =  \left\lag \; \left( \tlift {\cal O}_{i} (t)
                      - {\cal O}_{i} (t+1)  \tlift \right)
                      \; {\cal O}_{j} (0) \;
                 \right\rag
\lb{eq:cormat}
\eq
with $i,j=0,1,2,\ldots$ should decay proportional to $\exp(-W_{i} t)$.
Up to exponentially suppressed corrections this also holds
if the set of states is truncated at some value $i\!=\! r$ such
that $W_r$ is above the inelastic threshold. In this way we get
a system of independent states inside the elastic region \cite{luewo}.
In our simulations this requirement is fulfilled for $r\!\leq\! 6$
depending on $\kp$ and $L$.
%
\begin{figure}[tb]
\vbox to 0mm{%
     \centerline{
         \setlength{\unitlength}{7mm}
         \begin{picture}(13,10)(0.00,0.00)
         \put(1.00,7.20){\makebox(0.0,0.0)[bl]{\normalsize
         $\frac{E(\vct{k})}{E_{lat}(\vct{k})}$}}
         \put(1.00,3.00){\makebox(0.0,0.0)[bl]{\normalsize
         $\frac{E(\vct{k})}{\sqrt{m_\pi^2+\vct{k}^2}}$}}
         \put(7.80,-1.00){\makebox(0.0,0.0)[bc]{\normalsize $\vct{k}^2$}}
         \end{picture}
      }\vss}
  \centerline{
  \fpsxsize=70mm
  \fpsbox{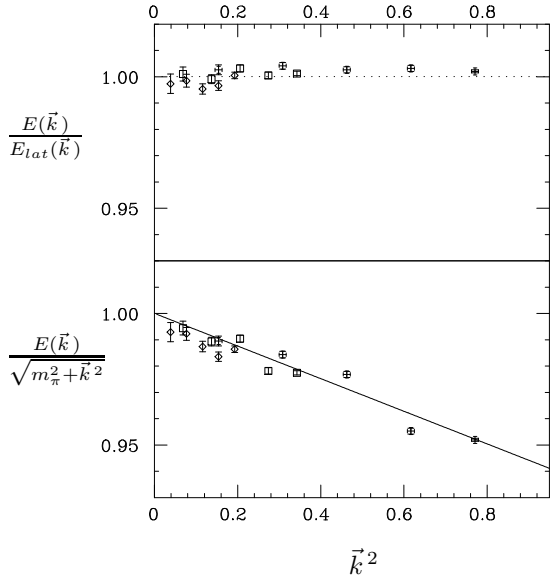}
  }
\caption{{\em Single particle energy values $E(\vct{k})$ divided by the
         lattice and the continuum dispersion relation, respectively, as
         function of $\vct{k}^2$} }
\vspace*{-4.0ex}
\lb{fig:disper}
\end{figure}
\vspace*{-0.7ex}
\section{TECHNICAL DETAILS}
\noindent
In order to generate our configurations
we used the reflection cluster algorithm in its multi-cluster version
generalised to the case of finite external source $J$. The Fourier transform
$\PP_{\vct{p},t}$ is computed by a 3-dimensional FFT. In the case of
the plane waves we also used the 2-cluster method \cite{luewo} to check our
results for the correlation matrix (\ref{eq:cormat}). The hopping parameter
$\kp=0.315$ and the external source $J=0.01$ are tuned such that the
$\sg$-mass lies in the elastic region.

The simulations are performed on cylindrical lattices $L^3\cdot T$.
On the HLRZ-CRAY Y-MP8/832 we can only run lattices of sizes
$16^3\cdot32$ and $20^3\cdot 32$.
The larger lattices $24^3 \cdot 32$, $32^3 \cdot 40$ are simulated
on the ``{\em Landesvektorrechner\/}'' SNI~S600/20 in Aachen.
In order to check the dependence on $T$ we have performed
runs on a lattice $24^3 \cdot 60$, but without getting significant changes
of the results.

Instead of calculating the eigenvalues of $C(t)$ we follow
refs.~\cite{luewo,gatlan}, choose some small $t_0$ and diagonalise
$C^{-1/2} (t_0) C (t) C^{-1/2} (t_0) $. After blocking the data we
determine the mean values and errors using the jackknife method. Finally
the energy values are extracted by a 1--parameter fit to the mean
eigenvalues according to the ansatz: $\exp(-(t-t_0)W_{i})$.

Variations of the rank $(r\! +\! 1)$ of the correlation function matrix have
only little influence on the results, while $t_0$ must be chosen small in
order to guarantee numerical stability of the eigenvalue calculation.

Furthermore we have developed a method to determine the upper end $t_{max}$
of the fit range dynamically.  The eigenvalues are sorted such that the
corresponding eigenvectors at successive time slices are as parallel as
possible. This should be equivalent to a sorting with respect to the size
of the eigenvalues, but due to statistical fluctuations at some time slice
$t=t_{max}\! +\! 1$ this fails, thus determining $t_{max}$.
\vspace*{-0.7ex}
\section{NUMERICAL RESULTS}
\noindent
In order to control scaling violations, we first examine the single
particle energies extracted from the exponential decay of the propagator
\be
    \left\lag\tlift\right.
    \sum_{a=1}^{3} \sum_{\vct{p}} \dl_{\,i,\vct{p}^2} \;
        \PP^{a}_{-\vct{p},t} \; \PP^{a}_{\vct{p},0}
    \left.\tlift\right\rag
        \sim e^{-E(\vct{k}) t} \,,
\eq
with $\vct{k}^2=(2\pi/L)^2\,i$. Fig.~\ref{fig:disper} shows that the data
are well approximated either by
\be
         E(\vct{k}) \approx \sqrt{m_\pi^2+\vct{k}^2}
                      \; \left(1 - 0.06(1) \, \vct{k}^2\,\right)
\eq
or by the lattice dispersion relation
($\hut{k}_{\nu} \!=\! 2 \sin\!\frac{k_{\nu}}{2}$):
\be
         E(\vct{k}) \approx E_{lat}(\vct{k})
         = 2 \asinh \left( \tlift \right. \frac{1}{2}
         \sqrt{m_\pi^2+\vct{\hut{k}}^2}
         \left. \tlift \right)
         \,.
\eq
Under the assumption that polarisation effects are exponentially suppressed
by a factor of $\exp(-m_\pi L)$ we get for the lowest state
($\vct{k}\!=\! \vct{0} $) the infinite volume mass $m_\pi=0.229(1)$.

Fig.~\ref{fig:woverl} shows our preliminary results for the two-particle
energy spectrum. The points  connected by solid lines are the numerical
results, bounded from above by the inelastic threshold $4m_\pi$.
%
\begin{figure}[tb]
\vbox to 0mm{%
     \centerline{
         \setlength{\unitlength}{6.5mm}
         \begin{picture}(13,10)(0.00,0.00)
         \put(0.40,8.00){\makebox(0.0,0.0)[bl]{
         ${{\ds W\vspace*{0.3ex}} \over {\ds m_\pi\tlift}}$}}
         \put(7.15,-0.60){\makebox(0.0,0.0)[bc]{$m_\pi L$}}
         \end{picture}
      }\vss}
  \centerline{
  \fpsxsize=65mm
  \fpsbox{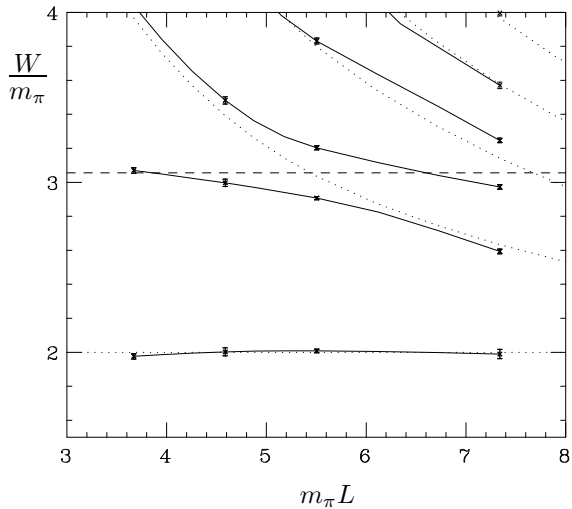}
  }
\vspace*{-1.5ex}
\caption{{\em Two-particle energy spectrum}}
\vspace*{-4.0ex}
\lb{fig:woverl}
\end{figure}
%
We get about the same numbers for both kinds of wave functions
(\ref{eq:plawav}), (\ref{eq:luewav}). The lowest value
represents the bound ground state, which is not relevant for the
discussion of scattering states.
The dotted lines correspond to the free energy spectrum:
\be
      \frac{W}{m_\pi} = 2 \sqrt{1 + \frac{\vct{k}^2}{m_\pi}} , \quad
              \vct{k} = \frac{2\pi}{L} \vct{n} , \quad
              \vct{n} \in \ZZ^{3} \,,
\eq
with $\vct{n}^2 = 0,1,2,\ldots$
We see the expected trend: the first and second level are coming close to
each other forming the so called ``avoided level crossing'',
such that we get an impression where the
resonance plateau (dashed line) might be.
%
\begin{figure}[tb]
\vbox to 0mm{%
     \centerline{
         \setlength{\unitlength}{6.5mm}
         \begin{picture}(13,10)(0.00,0.00)
         \put(0.60,8.20){\makebox(0.0,0.0)[bl]{\large $\dl_0$}}
         \put(6.85,-0.60){\makebox(0.0,0.0)[bc]{$k/m_\pi$}}
         \end{picture}
      }\vss}
  \centerline{
  \fpsxsize=65mm
  \fpsbox{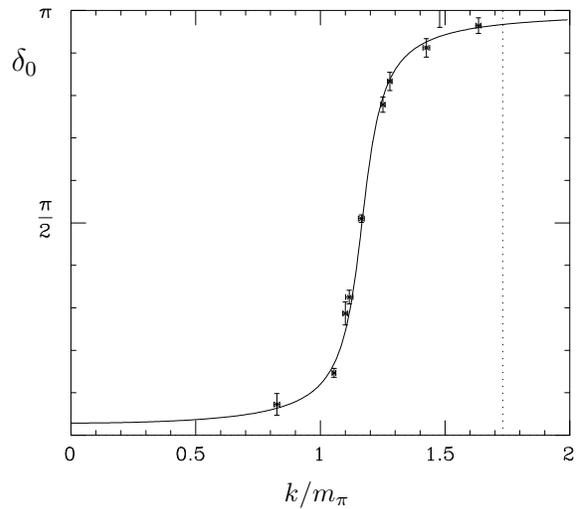}
  }
\vspace*{-1.5ex}
\caption{{\em Scattering phase shift $\dl_0$ in the isospin $0$ channel}}
\vspace*{-4.0ex}
\lb{fig:phase}
\end{figure}

For each lattice extent $L$ and energy level $W_{i}$ in the elastic region
the corresponding scattering phase shift is calculated according to
eq.~(\ref{eq:scatter}).
Fig.~\ref{fig:phase} summarises the results.
The vertical dotted line represents the inelastic threshold (see
eq.~(\ref{eq:elast})).

Finally we perform a fit to the data according to the Breit-Wigner
resonance formula (solid line)
\be
     \tan\left(\delta - \frac{\pi}{2}\right) = \frac{W-m_\sg}{\Gm_\sg/2}
\eq
determining the mass and the width of the $\sg$-resonance:
\be
     m_\sg   = 3.07(1) \; m_\pi \quad, \qquad
     \Gm_\sg = 0.19(2) \; m_\pi \,.
\lb{eq:masswid}
\eq
\vspace*{-0.7ex}
\section{DISCUSSION AND CONCLUSIONS}
\noindent
The resonance mass $m_\sg$ is about the same as the mass we have
obtained from a propagator fit.
Because of triviality, perturbation theory should be applicable in this
model, but the tree level approximation gives about three times the
numerical value for $\Gm_\sg$ in eq.~(\ref{eq:masswid}).
This deviation could at least partially be caused by lattice effects.
For further details the reader is referred to \cite{paper}.
\vspace*{-0.7ex}
\section*{ACKNOWLEDGEMENT}
\noindent
Helpful discussions with C.~Frick and M.~L\"uscher are gratefully
acknowledged. Special thanks are due to T.~Neuhaus for
allowing us to use his programs as basis of our programming.
Furthermore we wish to thank the {\em Rechenzentrum} at the RWTH Aachen
and the HLRZ J\"ulich for providing the necessary computer time.
\vspace*{-0.7ex}

\end{document}